\documentclass[twocolumn,showpacs,preprintnumbers,amsmath,amssymb,superscriptaddress,prb]{revtex4}
\usepackage{graphicx}
\usepackage{bm}

\begin{document}

\title{Observation of orientation- and $k$-dependent Zeeman spin-splitting
in hole quantum wires on (100)-oriented AlGaAs/GaAs
heterostructures}

\author{J. Chen}
\email{jchen@phys.unsw.edu.au} \affiliation{School of Physics,
University of New South Wales, Sydney NSW 2052, Australia}

\author{O. Klochan}
\affiliation{School of Physics, University of New South Wales,
Sydney NSW 2052, Australia}

\author{A.P. Micolich}
\email{mico@phys.unsw.edu.au} \affiliation{School of Physics,
University of New South Wales, Sydney NSW 2052, Australia}

\author{A.R. Hamilton}
\affiliation{School of Physics, University of New South Wales,
Sydney NSW 2052, Australia}

\author{T.P. Martin}
\affiliation{School of Physics, University of New South Wales,
Sydney NSW 2052, Australia}

\author{L.H. Ho}
\affiliation{School of Physics, University of New South Wales,
Sydney NSW 2052, Australia}

\author{U. Z\"{u}licke}
\affiliation{Institute of Fundamental Sciences and MacDiarmid
Institute for Advanced Materials \& Nanotechnology, Massey
University, Private Bag 11 222, Palmerston North 4442, New Zealand}

\author{D. Reuter}
\affiliation{Angewandte Festk\"{o}rperphysik, Ruhr-Universit\"{a}t
Bochum, D-44780 Bochum, Germany}

\author{A.D. Wieck}
\affiliation{Angewandte Festk\"{o}rperphysik, Ruhr-Universit\"{a}t
Bochum, D-44780 Bochum, Germany}

\date{\today}

\begin{abstract}
We study the Zeeman spin-splitting in hole quantum wires oriented
along the $[011]$ and $[01\overline{1}]$ crystallographic axes of a
high mobility undoped (100)-oriented AlGaAs/GaAs heterostructure.
Our data shows that the spin-splitting can be switched `on' (finite
$g^{*}$) or `off' (zero $g^{*}$) by rotating the field from a
parallel to a perpendicular orientation with respect to the wire,
and the properties of the wire are identical for the two
orientations with respect to the crystallographic axes. We also find
that the $g$-factor in the parallel orientation decreases as the
wire is narrowed. This is in contrast to electron quantum wires,
where the $g$-factor is enhanced by exchange effects as the wire is
narrowed. This is evidence for a $k$-dependent Zeeman splitting that
arises from the spin-$\frac{3}{2}$ nature of holes.
\end{abstract}
\maketitle

\section{Introduction}
The use of spin instead of charge to carry information is a central
goal in the fields of spintronics and quantum information,
generating significant interest in routes to efficient spin
manipulation in semiconductor devices~\cite{WolfSci01,ZuticRMP04}.
Low dimensional hole systems in p-type AlGaAs/GaAs heterostructures
hold considerable potential because the much stronger spin-orbit
coupling in holes~\cite{FertigSci03} may lead to devices where spin
can be manipulated electrostatically~\cite{DattaAPL90,
AwschalomNP07}.  The strong spin-orbit coupling also presents some
important fundamental physics questions, including how the peculiar
spin-$3/2$ nature of holes~\cite{WinklerBook03} is manifested in the
experimentally observable properties of low-dimensional GaAs hole
devices~\cite{WinklerPRL00, PapadakisPRL00, WinklerPRB04,
WinklerPRB05, DanneauAPL06, DanneauPRL06, CulcerPRL06, CsontosPRB07,
DanneauPRL08, KoduvayurPRL08, KlochanNJP09}.

Experiments to date have focussed almost solely on devices
fabricated in (311)-oriented AlGaAs/GaAs heterostructures. The
Zeeman spin-splitting in two-dimensional (2D) hole systems formed in
these heterostructures is highly anisotropic~\cite{WinklerPRL00,
PapadakisPRL00}, due to spin-orbit coupling and the low symmetry of
(311) surface. Recent studies have also revealed a significant
anisotropy in the Zeeman spin-splitting in one-dimensional (1D) hole
systems fabricated on the (311) heterostructures~\cite{DanneauPRL08,
KoduvayurPRL08, KlochanNJP09}, but it is not trivial to separate the
competing influences of 1D confinement and 2D crystallographic
anisotropy on the spin-splitting~\cite{KlochanNJP09}. Hole systems
fabricated on higher-symmetry planes such as (100) are not subject
to such complex crystallographic effects, and are therefore a much
better candidate for studying the spin physics of 1D hole systems.
To achieve high quality 1D hole systems we use
semiconductor-insulator-semiconductor field-effect transistor
(SISFET) devices, where a 2D hole system is `induced' using a
voltage applied to a degenerately-doped semiconductor gate rather
than through modulation doping~\cite{KaneAPL93,ClarkeJAP06}. Klochan
{\it et al.} have used this approach to fabricate 1D hole systems
with highly stable gate characteristics and clear conductance
quantization~\cite{KlochanAPL06}, and recently extended it to study
the Zeeman spin-splitting anisotropy in 1D hole systems in
(311)-oriented heterostructures~\cite{KlochanNJP09}.

In this paper, we extend this SISFET-based approach to study the
Zeeman spin-splitting in hole quantum wires oriented along the
$[011]$ and $[01\overline{1}]$ directions of a (100)-oriented
heterostructure. The crystallographic anisotropy that complicates
transport studies of quantum wires on (311)-oriented
heterostructures~\cite{KoduvayurPRL08, KlochanNJP09} does not occur
in these devices.  Instead, we find that the Zeeman spin-splitting
is finite when the applied magnetic field $B$ is oriented parallel
to the wire, and nearly zero when $B$ is oriented perpendicular to
the wire. This behaviour is almost identical for both orientations
of the wire relative to the dominant in-plane crystallographic
directions. The ability to switch the spin-splitting `on' or `off'
simply by rotating the applied magnetic field through $90^{\circ}$
may have useful spintronic applications. Finally, for $B$ parallel
to the wire, we observe $k$-dependent spin-splitting, where $g^{*}$
decreases as the wire is made narrower, in marked contrast to 1D
electron systems, where $g^{*}$ instead increases as the wire
becomes more one-dimensional~\cite{PatelPRB91A,ThomasPRL96}. This
finding is reminiscent of the absence of exchange enhancement
effects for 2D hole systems in (100)-oriented
heterostructures~\cite{WinklerPRB05}.

\section{Experimental Details}

Samples were fabricated from a (100)-oriented heterostructure that
consisted of a heavily doped $20$~nm C:GaAs cap, $10$~nm undoped
GaAs, an $160$~nm undoped AlGaAs barrier, and an undoped GaAs
buffer. The C-doped cap acts as a metallic gate~\cite{ClarkeJAP06},
with a 2D hole system induced at the AlGaAs/GaAs interface for
top-gate voltages $V_{TG} < -0.1$~V. Measurements of a separate
unpatterned Hall bar of the same heterostructure gave a peak
mobility $\mu = 4.8 \times 10^{5}$~cm$^{2}$/Vs at a density $p = 1.3
\times 10^{11}$~cm$^{-2}$ and a temperature $T = 100$~mK. The device
studied here consists of two orthogonal $400$~nm long quantum wires,
as shown in Fig.~1 (inset), defined by electron-beam lithography and
shallow wet etching of the cap layer. Each wire has three gates, a
top-gate used to control the density, and two side-gates used to
narrow the wire. The two wires can be measured independently and are
oriented along the $[011]$ and $[01\overline{1}]$ crystallographic
directions of a Hall bar running along the $[01\overline{1}]$
direction. The two wires are denoted as QW$011$ and
QW$01\overline{1}$, respectively. The quantum wires were measured in
a dilution refrigerator with a base temperature of $20$ mK using
standard a.c. lock-in techniques with an excitation voltage of
$50-100 ~\mu$V at a frequency of $17$~Hz. Measurements were obtained
at a top-gate voltage $V_{TG} = -0.80$~V, which corresponds to 2D
hole density of $2.56 \times 10^{11}$~cm$^{-2}$.

\begin{figure}
\includegraphics[width=8.5cm]{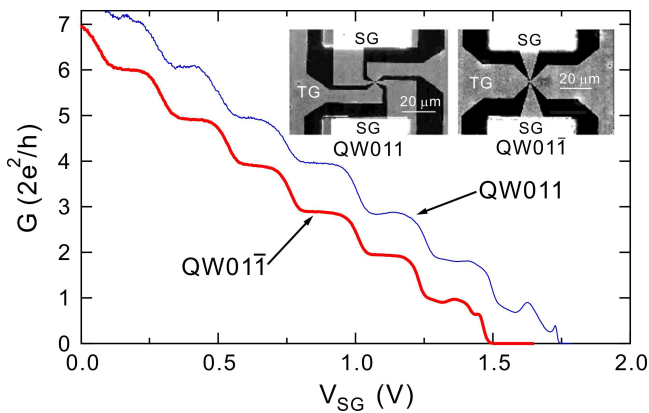}
\caption{The measured wire conductance $G$ versus side-gate voltage
$V_{SG}$ for QW$01\overline{1}$ (thick red line) and QW$011$ (thin
blue line).  The inset shows SEM micrographs of QW$011$ (left) and
QW$01\overline{1}$ (right), defined by electron-beam lithography
(EBL) and shallow wet-etching.}
\end{figure}

The width of the wire and its conductance $G$ can be gradually
reduced by applying a positive voltage $V_{SG}$ to the two
side-gates.  For both wires, we observe the well known `staircase'
of quantized conductance plateaus~\cite{VanWeesPRL88} as the wire is
narrowed by increasing $V_{SG}$, with the wire `pinching off' at
$V_{SG} \sim 1.5$~V. The similar pinch-off voltages indicate that
the two wires are almost identical, with similar dimensions and
confining potentials. The accurate quantization of the plateaus at
$G = n \times 2e^{2}/h$, where $n$ is the number of occupied 1D
subbands, confirms that transport through the wires is
ballistic~\cite{KlochanAPL06,VanWeesPRL88}. Moving from left to
right in Fig.~1 corresponds to strengthening the 1D confinement,
taking the wire from being only quasi-1D (large $n$ and $G$) towards
the 1D limit (small $n$ and $G$).

We study the spin properties of the hole quantum wires by measuring
the Zeeman spin-splitting for different orientations of the wire and
magnetic field with respect to the crystallographic axes.  To obtain
the $g$-factor for the various 1D subbands $n$, we use a technique
that compares the 1D subband splitting due to an in-plane magnetic
field~\cite{PatelPRB91} (see Fig.~2) and an applied d.c.
source-drain bias~\cite{GlazmanEPL89} (see Fig.~3). These two sets
of measurements are repeated in two cool-downs to allow for rotation
of the sample with respect to the magnetic field, thus providing
data for the four different combinations of wire and magnetic field
orientation with respect to the crystallographic axes.

\section{Results}
\subsection{1D subband spacings and source-drain bias measurements}

\begin{figure}
\includegraphics[width=8.5cm]{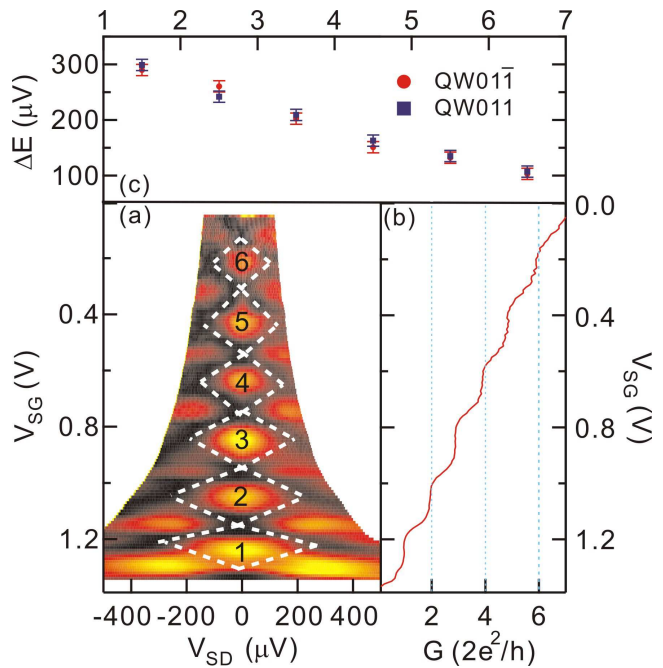}
\caption{(a) A colour-map of the transconductance $dg/dV_{SG}$
versus $V_{SD}$ on the $x$-axis and $V_{SG}$ on the $y$-axis for
QW$01\overline{1}$.  (b) The conductance $G$ vs $V_{SG}$ measured at
$V_{SD} = 0$~V, which corresponds to a vertical slice through the
center of the colour-map in (a).  The bright regions in (a)
correspond to conductance plateaus (low transconductance) and dark
regions correspond to the risers between plateaus (high
transconductance). The superimposed numbers in (a) indicate the
conductance $G$ of the corresponding plateau in (b) in units of
$2e^{2}/h$.  (c) The measured 1D subband energy spacings $\Delta
E_{n,n + 1}$ obtained from the subband crossings plotted as a
function of 1D subband index $n$.}
\end{figure}

The 1D subband spacing of the wires is obtained by adding a d.c.
bias $V_{SD}$ to the $20~\mu$V a.c. bias used to measure the
conductance. In Fig. 2(a) we plot the transconductance $dg/dV_{SG}$,
where $g = dI/dV$ is the differential conductance, as a colour-map
against $V_{SG}$ and $V_{SD}$ using data obtained from
QW$01\overline{1}$. Figure~2(b) shows the conductance $G$ vs
$V_{SG}$ measured at $V_{SD} = 0$~V and corresponds to taking a
vertical slice through the center of the colour-map in Fig.~2(a).
The dark regions in Fig.~2(a) correspond to high transconductance
(risers between plateaus) and the bright regions correspond to low
transconductance (the plateaus themselves).  Thus the dark regions
indicate when a particular 1D subband crosses the Fermi energy.  As
$V_{SD}$ is increased, the plateaus at multiples of $2e^{2}/h$
evolve into plateaus at odd multiples of $e^{2}/h$.  The subband
spacing $\Delta E_{n,n+1} = eV_{SD}$ is obtained from the
source-drain bias where adjacent transconductance peaks cross (i.e.,
from the dark regions at non-zero $V_{SD}$). The subband spacings
for the two wires are plotted in Fig.~2(c), and increase
monotonically from $\sim 100$ to $\sim 300~\mu$eV as the wire is
made narrower and more one-dimensional. The subband spacings for the
two wires agree to within $10~\mu$eV, again highlighting the
similarity of the two wires fabricated along different
crystallographic axes.

\subsection{Zeeman spin-splitting measurements}

The effect of an in-plane magnetic field $B$ on the 1D subbands is
shown in Fig.~3(a-d) for different orientations of the quantum wire
and magnetic field.  In each case we plot a colour-map of the
transconductance $dg/dV_{SG}$ versus $B$ and $V_{SG}$, with the dark
regions marking the 1D subband edges (high transconductance
corresponding to the risers between conductance plateaus). The
superimposed white dashed lines in Fig.~3 are guides to the eye
tracking the evolution of the 1D subbands with $B$.

\begin{figure*}
\includegraphics[width=17cm]{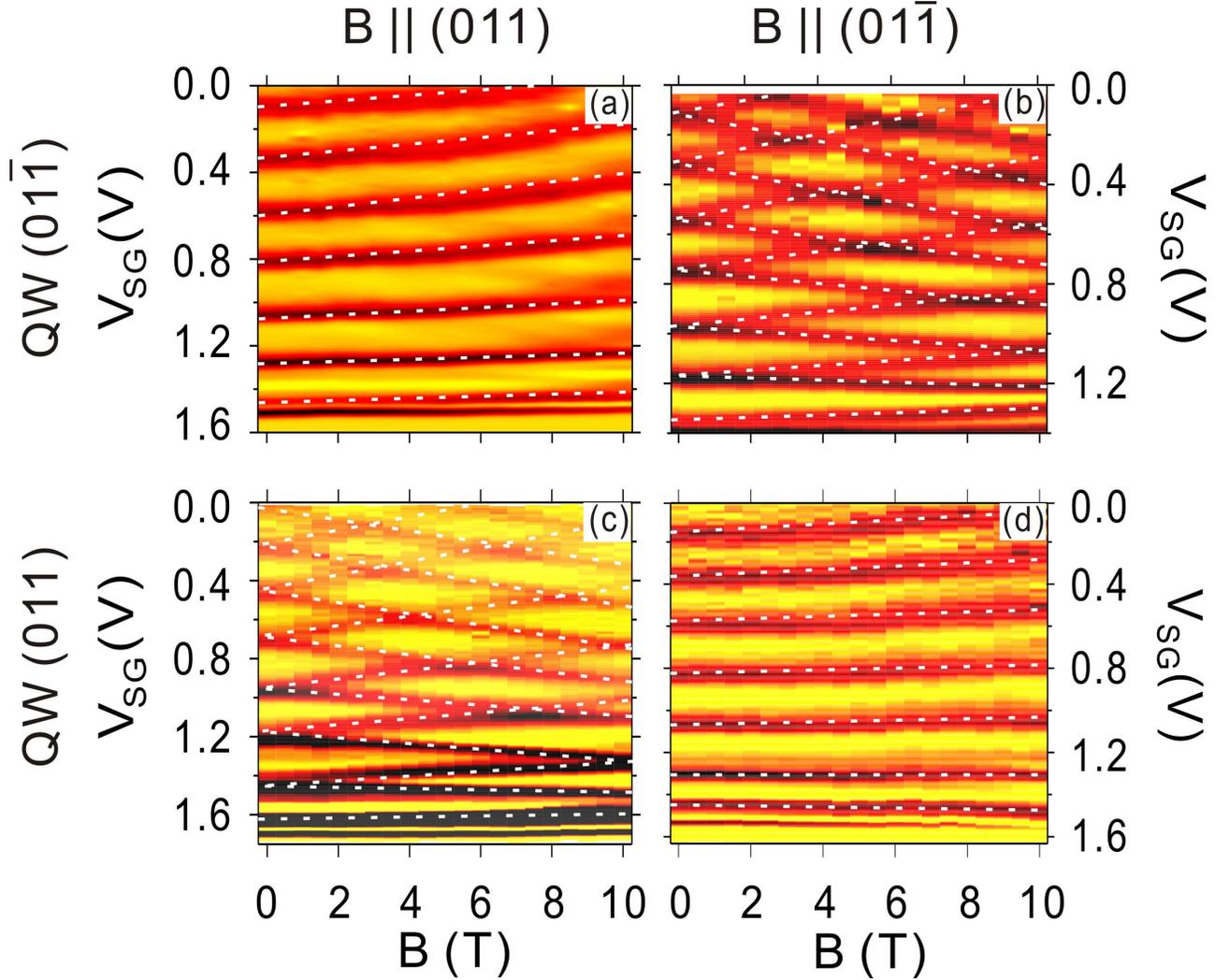}
\caption{Colour-maps of the transconductance $dg/dV_{SG}$ vs
in-plane magnetic field $B$ ($x$-axis) and $V_{SG}$ ($y$-axis) for
QW$01\overline{1}$ with (a) $B \parallel [011]$ and (b) $B
\parallel [01\overline{1}]$, and QW$011$ with (c) $B \parallel
[011]$ and (d) $B \parallel [01\overline{1}]$.  The superimposed
numbers in (a) indicate the conductance $G$ of the corresponding
plateau in units of $2e^{2}/h$. The white dashed lines are guides to
the eye that track the evolution of the various 1D subbands with
$B$.}
\end{figure*}

Figure 3 shows that there is only a Zeeman splitting of the 1D
subbands if $B$ is aligned along the wire, independent of the
crystallographic orientation of the wire: If the field is aligned
perpendicular to the wire, as in Figs.~3(a) and (d), then the Zeeman
spin-splitting is extremely weak. In Fig.~3(d) no splitting is
evident up to the highest fields available in the experiment $B =
10$~T, whilst in Fig.~3(a) some splitting is just apparent near $B
\sim 10$~T. In stark contrast, if $B$ is aligned parallel to the
wire, as in Figs.~3(b) and (c), then the Zeeman spin-splitting is
quite strong with clear splitting evident at quite modest fields $B
\sim 1$~T, crossings between adjacent subbands at moderate fields $B
\sim 5$~T, and ultimately, crossings between subbands differing in
$n$ by two at high fields $B \sim 10$~T. The directional-dependence
of the Zeeman spin-splitting in these (100)-oriented quantum wires
is much simpler than in wires fabricated on (311)-oriented
heterostructures, where a complex interplay between 1D confinement
and 2D crystallographic anisotropy is observed~\cite{DanneauPRL06,
KoduvayurPRL08, KlochanNJP09}.

\subsection{Obtaining the $g$-factors for the four magnetic field and
wire orientations}

We now extract the effective Land\'{e} $g$-factors~
\cite{DanneauPRL06, KlochanNJP09}. When $g^{*}$ is relatively large,
it can be obtained by measuring the field $B_{C}(n)$ at which the
spin down level of the $n^{th}$ subband crosses the spin-up level of
the $n + 1^{th}$ subband in Fig.~3. This crossing field, combined
with the corresponding d.c. bias $V_{SD}^{C}$ where the $n$ and $n +
1^{th}$ subbands cross in Fig.~2, gives:

\begin{equation}
\langle g^{*}_{n},g^{*}_{n + 1} \rangle =
\frac{eV_{SD}^{C}}{\mu_{B}B^{C}}
\end{equation}

\noindent Data obtained in this way are plotted as solid symbols at
$(n + 1)/2$ in Fig.~4, since they represent the average $g$-factor
for the two subbands.

\begin{figure}
\includegraphics[width=8.5cm]{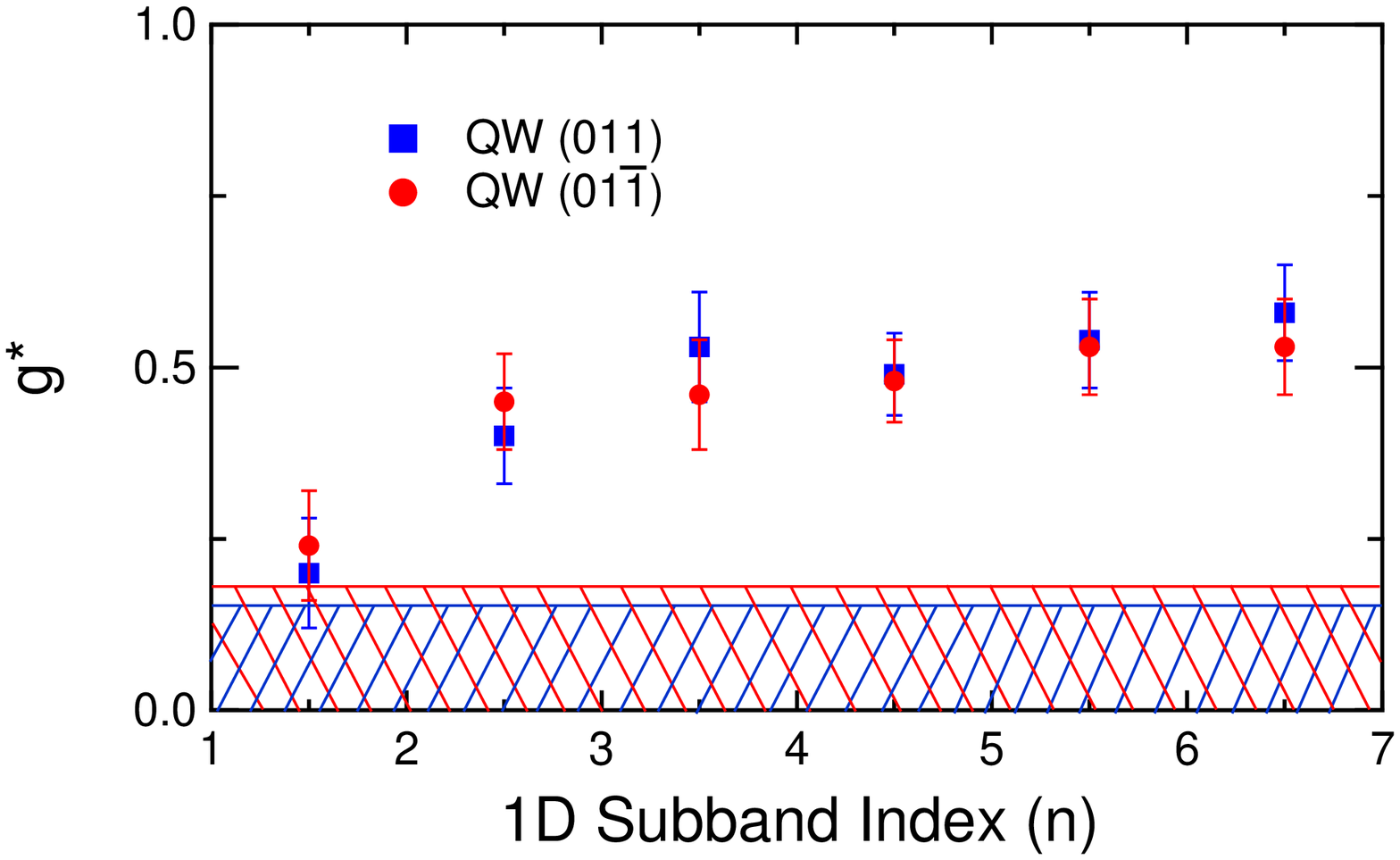}
\caption{The effective Land\'{e} $g$-factor $g^{*}_{n}$ plotted as a
function of 1D subband index $n$ for the four cases:
QW$01\overline{1}$ with $B \parallel [01\overline{1}]$ (red circles)
and $B \parallel [011]$ (blue squares); and QW$011$ with $B
\parallel [01\overline{1}]$ (red hatched region) and $B \parallel
[011]$ (blue hatched region). In the latter two cases, hatching is
presented because at best we can determine the upper bound on
$g^{*}$ as minimal spin-splitting is observed up to $B = 10$~T (see
text).}
\end{figure}

When the spin-splitting is small, as in Figs.~3(b) and (d), we can
only measure an upper bound on $g^{*}$ -- i.e., $g^{*}$ must sit
between zero and this upper bound otherwise the spin-splitting would
be resolvable. We determine this upper bound from the width $\Delta
V_{SG}$ of the transconductance peak in the colour-map at $B_{max} =
10$~T, which would be the maximum possible splitting if it could be
resolved. We convert this width into a splitting rate due to the
field $\partial V_{SG}/\partial B = \Delta V_{SG}/B_{max}$, and
combine it with d.c. biasing data $\partial V_{SG}/\partial V_{SD}$
to obtain the upper bound as:

\begin{equation}
|g^{*}| \leq \frac{e}{\mu_{B}} \frac{\partial V_{SD}}{\partial
V_{SG}} \frac{\Delta V_{SG}}{B_{max}}
\end{equation}

\noindent These upper bounds are indicated by the hatched regions in
Fig.~4.

\section{Discussion}
In order to discuss the two key results in Fig.~4, namely the
$g$-factor anisotropy and the decrease of $g^{*}$ as the wire is
made narrower, it is first necessary to review some of the
complexities of Zeeman splitting in the presence of spin-orbit
coupling.

For electrons in free space, an applied magnetic field causes the
spins to align along $\mathbf{B}$, with a spin splitting $\Delta E =
\pm \frac{1}{2} g \mu_B B$, with $g = 2$. In the presence of strong
spin-orbit coupling, the projections of spin and orbital angular
momenta $\mathbf{L}$ and $\mathbf{S}$ are no longer good quantum
numbers. Only the total angular momentum $\mathbf{J} = \mathbf{L} +
\mathbf{S}$ is conserved, and an applied magnetic field causes
$\mathbf{J}$ to align along $\mathbf{B}$. However if the electron is
in a non-symmetric environment, such as a polar GaAs crystal, a
quantum well, or a quantum wire, the quantisation axis for
$\mathbf{J}$ does not automatically align with the applied
$\mathbf{B}$ and it is rarely possible to find eigenstates of both
$\mathbf{B}$ and $\mathbf{J}$. This complicates the theoretical
analysis of spin splitting considerably, since the microscopic
details of the host crystal and the confinement due to the quantum
well/wire must all be taken into account.

\subsection{Zeeman splitting in 2D and quasi-1D holes}
The upper-most valence band in bulk GaAs consists of `heavy-hole'
(HH) and `light-hole' (LH) branches that are degenerate at the
valence band edge ($k = 0$). Confinement to a quantum well breaks
this HH-LH degeneracy, such that only the lowest HH subband ($m_{j}
= \pm \frac{3}{2}$) is occupied in a 2D hole system. However a
residual HH-LH coupling at finite wavevector not only results in
highly non-parabolic bands, but also plays a significant role in
determining the electronic properties in lower-dimensional hole
structures.

In the simplest approximation, the 2D confinement forces the
quantisation axis for $\mathbf{J}$ to point out of the 2D plane. To
lowest order there is only a spin-splitting of the HH states if $B$
is applied perpendicular to the quantum well, since $\langle
\mathbf{B.J} \rangle = 0$ for in-plane magnetic
fields~\cite{WinklerPRL00}. In practise however, the cubic
crystalline anisotropy terms~\cite{WinklerBook03}, as well as higher
order terms in the in-plane wave-vector $k_{\parallel}$, can result
in a finite in-plane Zeeman splitting. For quantum wells on the
(311) GaAs surface, the cubic anisotropy terms result in a linear in
$B_\parallel$ spin splitting at $k_{\parallel} = 0$. For (100)
oriented quantum wells the zeroth-order contributions due to cubic
crystalline anisotropies are absent~\cite{WinklerPRL00}, but a
substantial linear spin-splitting can still be achieved due to LH-HH
mixing at $\textbf{k}_{\parallel} \neq 0$, as discussed in \S7.4 of
Ref.~\cite{WinklerBook03}. Because the Zeeman splitting on (100)
surfaces arises from $k_{\parallel}$ dependent LH-HH mixing, it is
hard to define a $g$-factor for (100) 2D holes since $g^{*}$ must be
averaged over all occupied states, and is strongly dependent on
carrier density. One of the main advantages of quasi-1D systems
compared to 2D systems is the ability to perform energy
spectroscopy, and thereby measure the $g$-factor
directly~\cite{PatelPRB91A}.

We can predict the expected spin splitting in our quantum wire using
a quasi-1D model in which we take the 2D results in
Ref.~\cite{WinklerBook03} and add on the effects of quantisation of
the transverse wave-vector by the 1D confinement. We define the
components of the wavevector $\textbf{k}_{\parallel} =
(k_{l},k_{t})$ with respect to the axis of the quasi-1D wire, where
$k_{l}$ and $k_{t}$ are the in-plane wave-vector components parallel
(longitudinal) and perpendicular (transverse) to the quantum wire.
In the experiments on 1D holes, the spin-splitting is measured at
the 1D subband edges, where $k_{l} = 0$. Since $k_{t}$ is quantised
by the lateral 1D confinement, we can express the $g$-factor of the
$n^{th}$ 1D subband as:

\begin{eqnarray}
g^{*}_{[011],\parallel} (n) = g^{*}_{[01\overline{1}],\parallel}
(n) = 3\gamma_{3}|\kappa Z_{1} - 4\gamma_{3}Z_{2}|k_{n}^{2} \\
g^{*}_{[011],\perp} (n) = g^{*}_{[01\overline{1}],\perp} (n) =
3\gamma_{3}|\kappa Z_{1} - 4\gamma_{2}Z_{2}|k_{n}^{2}
\end{eqnarray}

\noindent Here $k_{n}$ is the quantised transverse wavevector
$k_{t}$ of the $n^{th}$ 1D subband, and we have rotated the
expressions in Eqn 7.22 of Ref.~\cite{WinklerBook03} to align along
the $[011]$ and $[01\overline{1}]$ axes. The subscripts on $g^{*}$
indicate the direction of the wire relative to the crystal and the
field relative to the wire, respectively. The terms $\gamma_{1}$,
$\gamma_{2}$ and $\gamma_{3}$ are Luttinger
parameters~\cite{LuttingerPR56} and $Z_{1,2}$ are LH-HH coupling
terms (see p. 147 of Ref.~\cite{WinklerBook03}).

The first inference we can draw from Eqns 3 and 4 is that the
$g$-factor for both $[011]$ and $[01\overline{1}]$ quantum wires
should exhibit the same anisotropy with respect to the magnetic
field, i.e., $g^{*}_{\parallel}/g^{*}_{\perp}$ is the same for both
wires. This is evident in Figs.~3 and 4. For both QW$011$ and
QW$01\overline{1}$, $g^{*}$ is the same for $B$ parallel to the wire
(see Fig.~3(b) and (c)) and very small for $B$ perpendicular to the
wire (see Fig.~3(a) and (d)). This behaviour is quite different to
quantum wires on (311) surfaces, where the anisotropy depends both
on the orientation of the quantum wire with respect to the magnetic
field and on the orientation of the field with respect to the
crystal axes~\cite{KlochanNJP09}.

However the quasi-1D theory disagrees with experiment on whether
$g^{*}_{\parallel} > g^{*}_{\perp}$ or $g^{*}_{\parallel} <
g^{*}_{\perp}$. Using expressions for $Z_{1,2}$ for square quantum
wells~\cite{WinklerBook03} and GaAs bandstructure parameters, Eqns~3
and 4 predict $g_{\perp} > g_{\parallel}$.  The experimental data
exhibits exactly the opposite trend, $g_{\perp} > g_{\parallel}$, as
shown in Fig.~4. This is a surprising result, and we have repeated
our experiment to confirm that this is indeed the case, obtaining
identical results (to within $10~\%$). We can only surmise that this
discrepancy lies in the dependence of $Z_{1,2}$ on the quantum-well
confinement, as our 2D holes are confined in a triangular potential
well at a single heterojunction, not in a square quantum well.
Unfortunately $Z_{1,2}$ are not available for a self-consistent
triangular quantum well.

A second conclusion we can draw from Eqns~3 and 4 is that in the
quasi-1D limit, the $g$-factor of the wires should decrease with
decreasing $k_{n}^{2}$. In the 1D constriction $k_{n}$ is given by
the difference between the Fermi energy in the 2D reservoirs
$E_{F}^{2D}$ and the bottom of of the 1D saddle-point
potential~\cite{ButtikerPRB90}. At large subband index $k_{n}$
approaches $k_{F}^{2D}$, and $g^{*}$ should saturate to a constant
value (as seen in Fig.~4). At small subband index the wire becomes
narrower, the saddle-point rises up in energy, $k_{n}$ decreases and
so does $g^{*}$. Additionally, the increase in 1D confinement
increases the LH-HH separation $\Delta E_{LH,HH}$, which reduces the
magnitude of the higher order Zeeman terms, and thereby reduces
$g^{*}$. The decrease in $g^{*}$ with decreasing subband index is
consistent with the data shown in Fig.~4, but is different to almost
all other studies of 1D systems, where a strong exchange enhancement
of $g^{*}$ is observed at low subband index~\cite{PatelPRB91A,
ThomasPRL96, DaneshvarPRB97, DanneauPRL06, MartinAPL08}. It is also
different to previous studies of 1D holes in (311) quantum wells,
where the Zeeman splitting is believed to be due to a combination of
crystal anisotropies at large $n$ and re-orientation of the
quantisation axis for $J$ at small $n$.

\subsection{Zeeman splitting in the 1D limit}

In the quasi-1D description, the 1D confinement is a weak
perturbation, so that $\mathbf{\hat{J}}$, the quantisation axis for
$J$, remains perpendicular to the 2D system. The lowest order terms
for the spin-splitting are zero, since $ \langle
\mathbf{B}.\mathbf{J} \rangle = 0$, and $g^{*}$ is only finite due
to the higher order $k_{\parallel}$ terms. It is thus interesting to
consider what happens in the 1D limit where the wire width becomes
equal to the width of the 2D confinement. In this case
$\mathbf{\hat{J}}$ is aligned with the wire axis and the lowest
order spin-splitting is large and positive for $B$ applied along the
wire, but is zero for $B$ perpendicular to the wire. This is
consistent with the anisotropy measured in Fig.~4, where
$g^{*}_{\parallel} > g^{*}_{\perp}$. If the 1D confinement is
causing a re-orientation of $\mathbf{\hat{J}}$, then one might
expect that $g^{*}$ would increase as the system is made more 1D, as
seen in previous experiments on (311) based hole
wires~\cite{DanneauPRL06,KlochanNJP09}. Furthermore, it is predicted
theoretically that the sign of $g^{*}$ is opposite for wires in the
$[011]$ and $[01\overline{1}]$ orientations for a square 2D
confinement, so one would expect the measured $g$-factors for the
two quantum wires to show different behaviour as we go from the
quasi-1D to the 1D limit.

Thus the quasi-1D model can explain the observed dependence of
$g^{*}$ on $k_{\parallel}$, but not the anisotropy of $g^{*}$,
whereas the 1D-limit model can explain the observed anisotropy of
$g^{*}$, but not the dependence on $k_{\parallel}$, since the latter
depends on the quasi-1D model. To be able to resolve this conundrum
it will be essential to perform more detailed calculations in the
quasi-1D limit for realistic 2D confining potentials.

\section{Conclusion}
In conclusion, we have studied the Zeeman spin-splitting in hole
quantum wires fabricated in (100)-oriented AlGaAs/GaAs
heterostructures, and find two new results: Firstly, if the applied
in-plane magnetic field $B$ is aligned along the wire, we see strong
spin-splitting, and if it is perpendicular to the wire, then we
observe negligible spin-splitting up to $B = 10$~T. This behaviour
is independent of the orientation of the wire on the heterostructure
surface. Although this latter finding is consistent with theoretical
predictions, our finding that the spin-splitting is maximized for
$B$ aligned along the wire is at odds with a quasi-1D theory, which
predicts maximum splitting instead for $B$ perpendicular to the
wire. At present the only solution to this disagreement may lie in
the sensitivity of the theoretical calculations on the 2D confining
potential -- theoretical results have only been obtained for a
square potential well so far, whereas the single heterojunction in
our device leads to a more triangular confinement. Secondly, we
report a decreasing $g^{*}$ as the 1D confinement is increased,
which is at odds with previous experiments of both 1D electron
systems in GaAs and InGaAs~\cite{PatelPRB91A, ThomasPRL96,
MartinAPL08}, and 1D hole systems in (311)-oriented GaAs
heterostructures~\cite{DanneauPRL06, DaneshvarPRB97}. This suggests
that despite the strong hole-hole interactions there is no exchange
enhancement in our 1D wires, consistent with recent measurements of
(100)-oriented 2D hole systems~\cite{WinklerPRB05}. These results
highlight the complex and interesting spin-physics associated with
$j = \frac{3}{2}$ hole systems, and suggest that much more
theoretical work is needed before we understand the physics of
holes, even on `simple' (100) surfaces.

\section{Acknowledgements}
This work was funded by Australian Research Council (ARC). ARH was
supported by an ARC Professorial Fellowship. DR and ADW thank the
DFG-SFB491, SPP1285 and the BMBF nanoQUIT for financial support. We
thank R. Winkler for helpful discussions and J. Cochrane for
technical support.

\end{document}